\documentclass[sigconf, screen]{acmart}

\acmConference[ICSE 2024]{46th International Conference on Software Engineering}{April 2024}{Lisbon, Portugal}

\setcopyright{none}
\renewcommand\footnotetextcopyrightpermission[1]{}
\AtBeginDocument{%
  \providecommand\BibTeX{{%
    \normalfont B\kern-0.5em{\scshape i\kern-0.25em b}\kern-0.8em\TeX}}}

\usepackage{hyperref}
\usepackage{boldline}
\usepackage{color,colortbl}
\usepackage{bigstrut}
\usepackage[ruled]{algorithm2e}
\usepackage{amsmath,amsfonts}
\usepackage{amsmath}
\usepackage{array}
\usepackage{algorithmic}
\usepackage{balance}
\usepackage{blindtext}
\usepackage{booktabs}
\usepackage{hyperref}
\usepackage[noabbrev]{cleveref}
\usepackage{comment}
\usepackage{enumitem}
\usepackage{eqparbox}
\usepackage{fancybox}
\usepackage{fancyvrb}
\usepackage{framed}
\usepackage{graphicx}
\usepackage{ifthen}
\usepackage{listings}
\usepackage{mathrsfs}
\usepackage{mdwmath}
\usepackage{mdwtab}
\usepackage{multirow}
\usepackage{pifont}
\usepackage{textcomp}
\usepackage{url}
\usepackage{xspace}
\usepackage[normalem]{ulem}
\usepackage[framemethod=TikZ]{mdframed}
\usepackage{caption}
\usepackage{subcaption}
\usepackage{tabularx}
\usepackage{tcolorbox}
\tcbuselibrary{listings,skins}
\usepackage{mathtools}
\usepackage{blindtext}
\usepackage{lstautogobble}
\usepackage{colortbl}

\DeclareGraphicsExtensions{.pdf,.jpeg,.png}
\graphicspath{{figures/}}

\newcommand{\rom}[1]{\uppercase\expandafter{\romannumeral #1\relax}}

\newcommand{\etal}{\hbox{\emph{et al.}}\xspace}
\newcommand{\eg}{\hbox{\emph{e.g.,}}\xspace}
\newcommand{\ie}{\hbox{\emph{i.e.,}}\xspace}

\newcommand{\wrt}{\hbox{\emph{w.r.t.}}\xspace}

\newcommand{\vs}{\hbox{\emph{vs.}}\xspace}
\newcommand{\etc}{\hbox{\emph{etc.}}\xspace}

\newlength\Linewidth
\def\findlength{\setlength\Linewidth\linewidth
\addtolength\Linewidth{-4\fboxrule}
\addtolength\Linewidth{-3\fboxsep}
}
\newenvironment{examplebox}{\par\begingroup
  \setlength{\fboxsep}{3pt}\findlength
  \setbox0=\vbox\bgroup\noindent
  \hsize=0.90\linewidth
  \begin{minipage}{0.90\linewidth}\normalsize}
    {\end{minipage}\egroup
    \textcolor{gray20}{\fboxsep1.2pt\fbox
     {\fboxsep5pt\colorbox{gray05}{\normalcolor\box0}}}
    \endgroup\par\noindent
    \normalcolor\ignorespacesafterend}

\usetikzlibrary{shadows}
\usepackage{graphics}
\newmdenv[
    tikzsetting= {fill=blueish},
    skipabove=0.33em,
    skipbelow=0.33em,
    linewidth=1pt,
    innerleftmargin=4pt,
    innerrightmargin=4pt,
    innertopmargin=2pt,
    innerbottommargin=2pt,
    linecolor=gray95,
    roundcorner=2pt, 
    shadow=true,
    shadowsize=4pt,
    shadowcolor=gray95
]{questionbox}

\newmdenv[
    tikzsetting= {fill=greenish},
    skipabove=0.33em,
    skipbelow=0.33em,
    linewidth=1pt,
    innerleftmargin=4pt,
    innerrightmargin=4pt,
    innertopmargin=2pt,
    innerbottommargin=2pt,
    linecolor=gray95,
    roundcorner=2pt, 
    shadow=true,
    shadowsize=4pt,
    shadowcolor=gray95
]{answerbox}

\newmdenv[
    skipabove=0.33em,
    skipbelow=0.33em,
    innerleftmargin=4pt,
    innerrightmargin=4pt,
    innertopmargin=2pt,
    innerbottommargin=2pt,
]{lessonbox}

\usepackage{tikz}
\definecolor{javared}{rgb}{0.6,0,0} % for strings
\definecolor{javagreen}{rgb}{0.25,0.5,0.35} % comments
\definecolor{javapurple}{rgb}{0.5,0,0.35} % keywords
\definecolor{javadocblue}{rgb}{0.25,0.35,0.75} % javadoc

\lstdefinestyle{basejava}{
  language=java,
  showstringspaces=false,
  basicstyle=\small\ttfamily,
  keywordstyle=\bfseries\color{javapurple},
  commentstyle=\itshape\blue,
  identifierstyle=\blue,
  frame=none,
  backgroundcolor=\color{white},
}

\lstdefinestyle{CustomJava}{
  belowcaptionskip=\baselineskip,
  breaklines=true,
  xleftmargin=\parindent,
  language=java,
  showstringspaces=false,
  basicstyle=\scriptsize\ttfamily,
  keywordstyle=\bfseries\color{javapurple},
  commentstyle=\itshape\blue,
  identifierstyle=\blue,
  belowskip=1pt,
  numbers=left,
  gobble=0
}

\lstdefinestyle{CustomJavaWoNumbers}{
  belowcaptionskip=0.5\baselineskip,
  breaklines=true,
  xleftmargin=\parindent,
  language=java,
  showstringspaces=false,
  basicstyle=\scriptsize\ttfamily,
  keywordstyle=\bfseries\color{javapurple},
  commentstyle=\itshape\blue,
  identifierstyle=\blue,
  belowskip=0.5pt,
  numbers=none,
  gobble=0
}

\lstdefinestyle{codit}{
  belowcaptionskip=\baselineskip,
  breaklines=true,
  language=java,
  showstringspaces=false,
  basicstyle=\scriptsize\ttfamily,
  keywordstyle=\bfseries\color{javapurple},
  commentstyle=\itshape\blue,
  identifierstyle=\blue,
}

\newcommand\blue[1]{\textcolor[rgb]{0.00,0.00,1.00}{{#1}}}

\newcommand\dkgreen[1]{\textcolor[rgb]{0.0,0.6,0}{\textbf{#1}}}

\definecolor{blueish}{RGB}{250, 250, 255}
\definecolor{greenish}{RGB}{250, 255, 250}
\definecolor{redish}{RGB}{255, 200, 200}

\definecolor{highlight}{RGB}{175, 255, 100}
\definecolor{gray01}{gray}{.98}
\definecolor{gray05}{gray}{0.95}
\definecolor{gray08}{gray}{0.92}
\definecolor{gray10}{gray}{0.90}
\definecolor{gray12}{gray}{0.88}
\definecolor{gray15}{gray}{0.85}
\definecolor{gray18}{gray}{0.82}
\definecolor{gray20}{gray}{0.80}
\definecolor{gray25}{gray}{0.75}
\definecolor{gray30}{gray}{0.70}
\definecolor{gray35}{gray}{0.65}
\definecolor{gray40}{gray}{0.60}
\definecolor{gray45}{gray}{0.55}
\definecolor{gray50}{gray}{0.50}
\definecolor{gray55}{gray}{0.45}
\definecolor{gray60}{gray}{0.40}
\definecolor{gray65}{gray}{0.35}
\definecolor{gray70}{gray}{0.30}
\definecolor{gray75}{gray}{0.25}
\definecolor{gray80}{gray}{0.20}
\definecolor{gray85}{gray}{0.15}
\definecolor{gray90}{gray}{0.10}
\definecolor{gray95}{gray}{0.05}
\definecolor{rowgray}{RGB}{224, 224, 224}
\newcommand{\tool}{\textsc{TRACED}\xspace}

\newtcbox{\inlinebox}[1][]{enhanced,
 box align=base,
 nobeforeafter,
 colback=blueish,
 size=small,
 left=0pt,
 right=0pt,
 boxsep=2pt,
 #1}

\renewcommand{\cref}[1]{\Cref{#1}}

\usepackage{tcolorbox}
\newcounter{findingCounter}
\newenvironment{finding}{
\begin{tcolorbox}[colback=blue!5!white,colframe=blue!5!white,arc=0mm,grow to left by=0mm,left=0mm,grow to right by=0mm,left=1.5mm,right=1.5mm,top=1.5mm,bottom=1.5mm]
\textbf{Result-\arabic{findingCounter}\stepcounter{findingCounter}:}
}
{
\end{tcolorbox}
}

\lstdefinelanguage{c-pretty}
{
  language=c,
  numbers=left,
  basicstyle=\footnotesize\ttfamily,
  numberstyle=\footnotesize,
  breaklines=true,
  columns=fullflexible,
  xleftmargin=0pt,
  showstringspaces=false,
  identifierstyle=\color{black},
  keywordstyle=\color{javapurple}\bfseries,
  stringstyle=\color{javared},
  commentstyle=\color{javagreen},
  morecomment=[s][\color{javadocblue}]{/**}{*/},
}

\setcopyright{acmcopyright}
\copyrightyear{2018}
\acmYear{2018}
\acmDOI{XXXXXXX.XXXXXXX}

\acmPrice{15.00}
\acmISBN{978-1-4503-XXXX-X/18/06}

\begin{document}

\title{TRACED: Execution-aware Pre-training for Source Code}

\author{Yangruibo Ding}
\affiliation{%
  \institution{Columbia University}
  \city{New York}
  \state{NY}
  \country{USA}
}
\author{Ben Steenhoek}
\affiliation{%
    \institution{Iowa State University}
    \city{Ames}
  \state{IA}
  \country{USA}
}

\author{Kexin Pei}
\affiliation{%
    \institution{Columbia University}
    \city{New York}
    \state{NY}
    \country{USA}
}

\author{Gail Kaiser}
\affiliation{%
    \institution{Columbia University}
    \city{New York}
  \state{NY}
  \country{USA}
}

\author{Wei Le}
\affiliation{%
    \institution{Iowa State University}
    \city{Ames}
  \state{IA}
  \country{USA}
}

\author{Baishakhi Ray}
\affiliation{%
    \institution{Columbia University}
    \city{New York}
  \state{NY}
  \country{USA}
}

\begin{abstract}
Most existing pre-trained language models for source code focus on learning the static code text, typically augmented with static code structures (abstract syntax tree, dependency graphs, \etc). However, program semantics will not be fully exposed before the real execution. Without an understanding of the program execution, statically pre-trained models fail to comprehensively capture the dynamic code properties, such as the branch coverage and the runtime variable values, and they are consequently less effective at code understanding tasks, such as retrieving semantic clones and detecting software vulnerabilities.

To close the gap between the static nature of language models and the dynamic characteristics of programs, we introduce \tool, an execution-aware pre-training strategy for source code. Specifically, we pre-train code language models with a combination of source code, executable inputs, and corresponding execution traces. Our goal is to teach code models the complicated execution logic during the pre-training, enabling the model to statically \emph{estimate} the dynamic code properties without repeatedly executing code during task-specific fine-tuning.

To illustrate the effectiveness of our proposed approach, we fine-tune and evaluate \tool on three downstream tasks: static execution estimation, clone retrieval, and vulnerability detection. The empirical results show that \tool relatively improves the statically pre-trained code models by 12.4\% for complete execution path prediction and by 25.2\% for runtime variable value predictions. \tool also significantly outperforms statically pre-trained models in clone retrieval and vulnerability detection across four public benchmarks.

\end{abstract}

\maketitle
\section{Introduction}

\label{sec:intro}

Machine Learning (ML) for source code 
has enabled many software engineering tasks, such as automated program repair~\cite{ding2020patching, jiang2021cure, jiang2023impact, jiang2023knod}, 
bug finding~\cite{chakraborty2021reveal, Zhou2019DevignEV}, and refactoring~\cite{chakraborty2021natgen}.
Recently, the common practice of training ML models for source code understanding is based on pre-training a Transformer-based language model on source code.
These approaches treat source code programs as \emph{static text}~\cite{feng2020codebert, buratti2020cbert, ahmad2021plbart, wang2021codet5}, sometimes augmented with program-specific structures such as abstract syntax trees and dependency graphs~\cite{guo2021graphcodebert, guo2022unixcoder, ding2021disco, niu2022sptcode}, and adapt pre-training strategies for natural language to learn program representations.

However, many source code understanding tasks require a more comprehensive understanding of \emph{program behavior}. 
For instance, detecting semantic clones\cite{msr2021codexglue} involves determining if two pieces of code behave similarly under similar inputs, even if their structures are apparently different. Likewise, detecting vulnerabilities often requires developers to analyze whether a potentially problematic location can be executed and what kinds of value flows can expose any vulnerability. While existing code models are primarily trained to capture static code properties, they are not effective at reasoning about program behavior. In fact, many of the deeper program semantics only manifest when the code is executed. As a result, they tend to underperform when it comes to tasks that require deeper semantic understanding.

\begin{figure}[!th]
    \centering
    \includegraphics[width=\columnwidth]{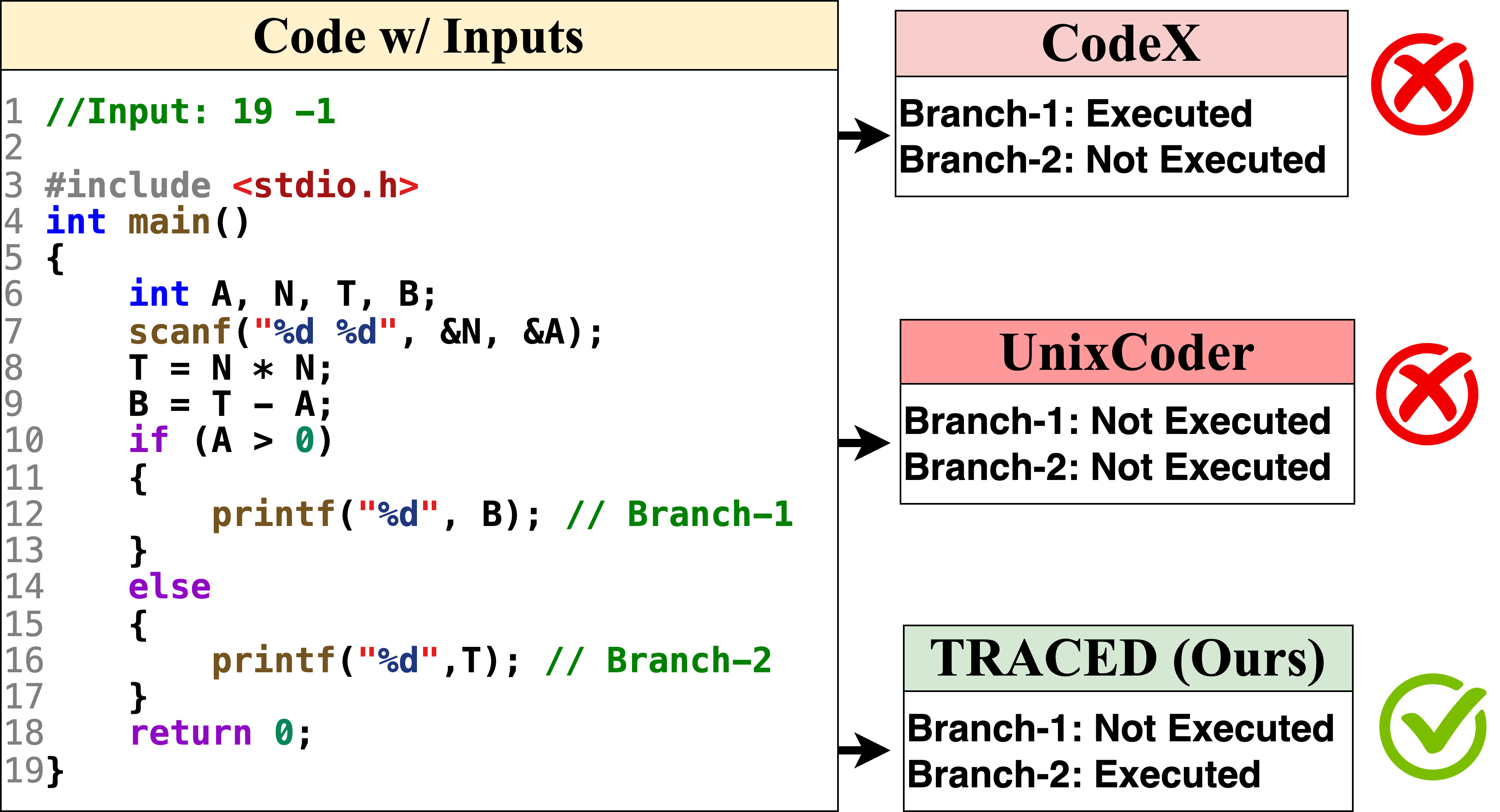}
    \caption{\small An motivating example from CodeNet's coding challenge No.3597~\cite{puri2021codenet} reveals that statically pre-trained code language models, regardless of their size, could not reason about the branch coverage given a specific input, while \tool, enhanced with program execution features, correctly identify the execution path.}
    \label{fig:motivation}
    %\vspace{-5mm}
\end{figure}

\vspace{-2mm}
\paragraph{\textbf{Motivating Examples.}} Figure~\ref{fig:motivation} presents an example with simple execution logic to illustrate the failure of statically pre-trained code models on the branch coverage prediction. 
We query three pre-trained code models, CodeX~\cite{chen2021codex} (code-davinci-002), 
UnixCoder~\cite{guo2022unixcoder}, and \tool (ours), to predict the branch coverage, according to the given program inputs. 
For CodeX, we prompt the model with carefully designed questions, similar to~\cite{nye_show_2021}, to ask for the branch coverage prediction in the zero-shot setting. 
Specifically, we augment the prompts by adding comments at the end of lines 12 and 16: \texttt{// Will this line be executed? Yes or no?}. 
To give more hints regarding the data flow, we further add a comment at the end of line 10: \texttt{// A is -1, since it accepts the second value of the input}. 
Unfortunately, even if provided with additional hints of the required data flow for branch prediction, CodeX still failed to predict the correct coverage labels, suggesting it cannot interpret this simple execution.  

Besides the zero-shot prompting, we also study whether fine-tuning pre-trained code models to predict execution can lead to better branch prediction. 
Specifically, we fine-tune another popular pre-trained code model, UnixCoder~\cite{guo2022unixcoder}, to predict branch execution while ensuring the motivation example is not seen during training. 
From the inference results in Figure~\ref{fig:motivation}, we notice that UnixCoder cannot predict covered branches even after being fine-tuned. 
It predicts neither of the branches will be covered, indicating that it does not have the basic understanding that, for this specific example, at least one branch will always be taken on a valid input.

\paragraph{\textbf{Our approach.}} To address the limitation of the statically pre-trained code models, we propose \tool, an execution-aware pre-training strategy to capture the static and dynamic perspectives of the source code. 
Specifically, we pre-train the Transformer-based language model with multi-task objectives on predicting source code, program states, and execution coverage, forcing the model to reason about both program's runtime behavior and the naturalness of the source code~\cite{ray2016naturalness} at the same time.
We address several technical challenges, such as representing program execution states, encoding the runtime variable values, and representing code coverage, to implement the pre-training strategy.

\paragraph{Representing Program States.} During program execution, variables are used to store data that is used by the program. These variables can have different types, such as integers, floating-point numbers, pointers, and arrays. As the program executes, the values of these variables change, reflecting the changes in the program's state. Consequently, software developers typically monitor the variable values, via debugging tools, to observe the execution facts~\cite{zeller2005why} and understand the dynamic behaviors of the program. 

In this work, we define the \emph{program state} at a specific time step of the execution as the set of values of every defined variable in the current scope. In other words, the program state is equivalent to the value mapping table of the debugger, which is monitored by the developer when the program is paused by a specific breakpoint.

%\vspace{-2mm}
\paragraph{Value Quantization.} While the runtime variable values are traced as concrete values, directly learning them brought challenges to machine learning models. Concrete values span over a wide range of possible values, especially when considering different data types (integers, floating-point numbers, arrays, pointers, etc.), leading to a high-dimensional, complex, but sparse data distribution. % \bent{which the code model must learn to fit}. 
This increased data complexity and sparsity challenges the model to learn patterns and relationships between the variable values, as it must deal with many unique inputs, which causes the model to overfit and memorize specific instances rather than generalize to broader patterns. Additionally, noise, outliers, and irregularities of concrete values also mislead the model's learning process. We will empirically demonstrate these limitations in \S\ref{subsec: rq_quantized_value}.

To decrease the data complexity and increase the density, we define thirty value categories, covering a wide range of variable types, to map the continuous but sparse variable values into discrete bins. We call this process as \emph{value quantization}, which is similar in design to the quantization in signal processing\footnote{\url{https://en.wikipedia.org/wiki/Quantization_(signal_processing)}}. This simplification potentially helps the model to be more resilient to noise and outliers, allowing it to focus on learning the underlying execution patterns and relationships between variables, rather than being sensitive to specific instances or irregularities. 

\paragraph{Representing Execution Coverage.} While program state labels provide important information about the current state of the program, they do not capture information about how the program arrived at that state. To boost the training with more comprehensive execution features, besides the variable values, we also log the execution coverage during the execution, in terms of which lines are executed and which are not, and construct execution coverage features for the model to learn.

\paragraph{\textbf{Results.}} We fine-tune and evaluate \tool's performance using three tasks: static execution estimation, clone retrieval, and vulnerability detection.
On statically predicting the program executions, \tool substantially improves the statically pre-trained code models by 12.4\% for execution path prediction and by 25.2\% for runtime variable value predictions.
\tool also obtains state-of-the-art results in code understanding tasks: \tool reports 91.2\% MAP@R on CodeXGLUE-POJ104~\cite{msr2021codexglue}, 50.4\% F1 on ReVeal~\cite{chakraborty2021reveal}, and 65.9\% accuracy on CodeXGLUE-defect-detection~\cite{msr2021codexglue}.

\paragraph{\textbf{Contributions.}}
We make the following contributions:
\begin{itemize}
\item We present a simplified and compact representation of program executions, including the program states and the execution coverage, to effectively guide code models to learn program semantics and reason about program behavior.

\item We propose a novel multi-task pre-training strategy to jointly learn the static and dynamic code properties. As a result, the pre-trained model with our approach will be empowered with a decent execution awareness.

\item We pre-train \tool with the proposed trace representation and the execution-aware strategy and evaluate its performance on several downstream tasks. The experiment results demonstrate that \tool significantly outperforms the statically pre-trained code models in these tasks.

\item We will publicly release our data, code, and pre-trained models 
to foster open science.
\end{itemize}
\section{Overview}
\label{sec:overview}
\begin{figure}[th]
    \centering
    \includegraphics[width=\columnwidth]{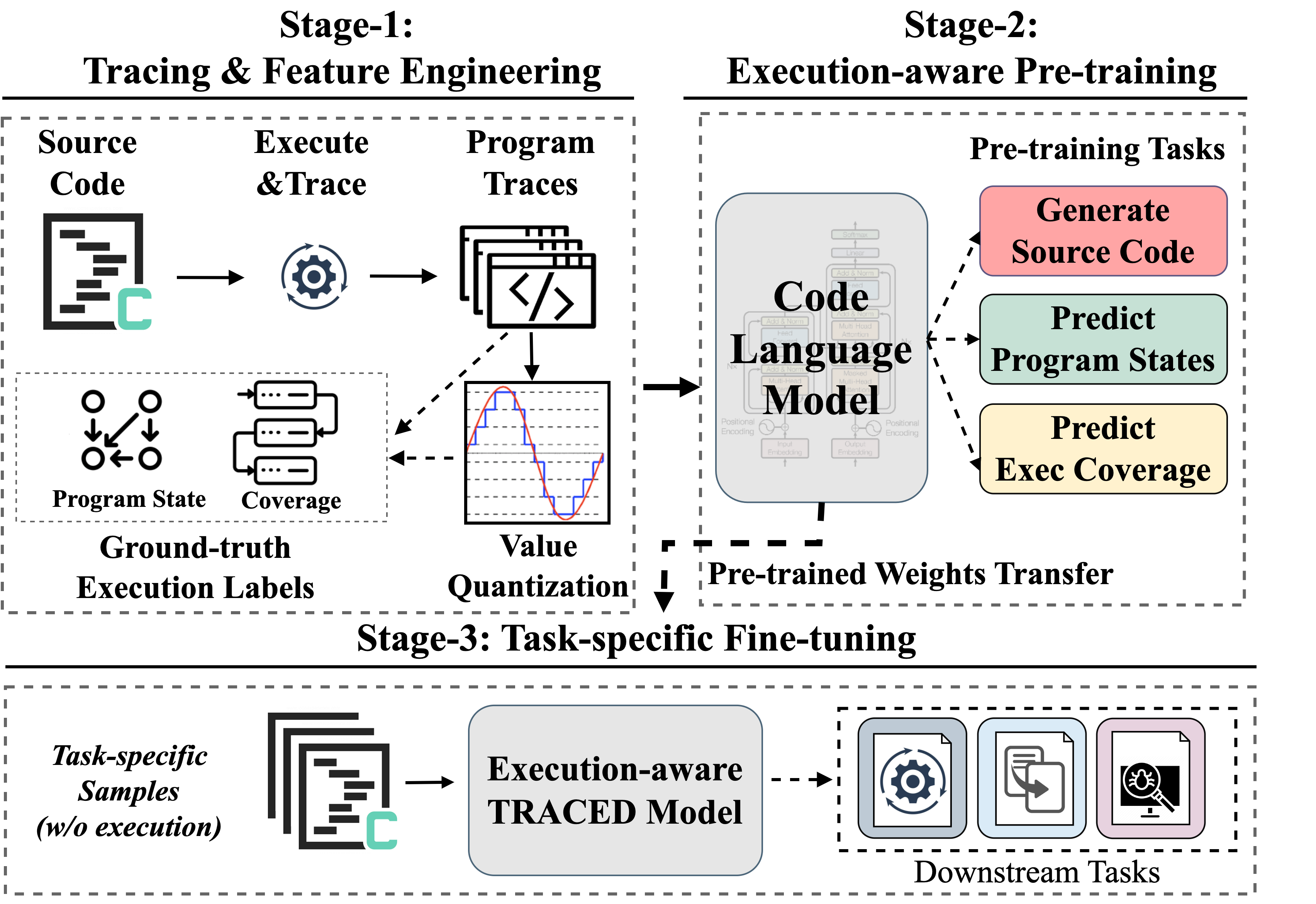}
    \caption{\small 
    Overview of \tool workflow.
    }
    \label{fig:overview}
\end{figure}

Figure~\ref{fig:overview} shows the overview of \tool, consisting of three main stages: (1) tracing the source code and engineering the features, (2) execution-aware pre-training using the program traces, and (3) loading the pre-trained weights and performing task-specific fine-tuning.

\paragraph{\textbf{Stage-1: Tracing \& Feature Engineering.}} The goal of this stage is to prepare the data for pre-training. The process begins with providing a source program and its executable inputs. The first step is to execute the program with each input to generate corresponding traces. The traces record the runtime variable values, together with the execution coverage, logging the full execution history of the program and revealing the changes to program states throughout execution. To reduce the complexity and sparsity of the data, and make it easier for the model to learn patterns and relationships between the variable values, we quantize the concrete runtime values recorded in the traces into pre-defined value ranges. The quantization process maps continuous values to a fixed set of discrete or bins. By quantizing the values, we create a finite set of possible outputs that can be used as ground-truth labels during training. After quantization, we create program state labels and execution coverage labels that will help the model to capture the program executions. The dataset finally ends up with a set of samples and labels, where each sample includes the source code with its program input and the labels represent the execution trace of this sample.

\paragraph{\textbf{Stage-2: Execution-aware Pre-training with Traces.}} We utilize the pre-processed samples and labels obtained from Stage-1 to perform supervised pre-training. Specifically, we use a Transformer-encoder-based model~\cite{liu2019roberta} to learn the program traces and improve the model's understanding of program execution. The model could be either trained from scratch or loaded by the pre-trained weights of existing code language models. To achieve the goal of producing execution-aware code representation, we propose three pre-training objectives. The first objective is learning to generate the source code. We believe that understanding the naturalness of code text~\cite{hindle2012naturalness, ray2016naturalness} is fundamental for the model to capture more sophisticated signals such as program execution. This objective is implemented with masked language modeling (MLM), which masks a certain percentage of tokens in the source code and trains the model to reconstruct the masked tokens based on the surrounding context. The second objective is learning to predict the program states. By predicting program state labels that were generated in Stage-1, the model learns to capture the data flows and the side effects of code execution. The third objective is to predict the execution coverage. By predicting the execution coverage labels generated by Stage-1, the model learns to capture the dynamic control flow and helps the model understand how the program state is reached and evolving.

\paragraph{\textbf{Stage-3: Task-specific Fine-tuning.}} Finally, we apply \tool to several downstream tasks. We load the pre-trained weights of \tool, fine-tune the model for a specific task, and keep updating the model weights. Fine-tuning does not require the program to be executed; rather, \tool will reason about the execution statically with its learned execution signals during the pre-training, and learn to accomplish the task accordingly. In many useful applications, we would not have program traces available. We consider three downstream tasks for \tool: static execution estimation, which includes execution coverage and runtime variable value predictions, clone retrieval, and vulnerability detection.

\section{Tracing \& Feature Engineering}
\label{sec: tracing_feature}

%\gail{this section talks about how "we" do something and not how \tool does something, shouldn't we mention \tool in every section?}

In this section, we introduce how \tool builds the learnable features from program traces for models to learn the program executions.

\subsection{Representing Program States} 
\label{subsec:program_state}

To imitate the way that human developers monitor variable values to understand program behavior, we propose to train neural models with the log of runtime variable values to recognize execution patterns and infer dynamic program behaviors in a way that is similar to human intuition. By taking the log of variable values during the execution, we can represent the program states in a more compact and interpretable form that is manageable for deep neural nets to process.

We build the program state by taking snapshots of variable values at program points during execution. When we take a snapshot at a specific time step, similar to the moment that the program is paused by a debugging breakpoint set right after line $l$, we maintain a value mapping, $M$, to map the variable to its current value, similar to the value mapping table of the debugger. To record the program state, we take the value snapshot after
%\gail{should this say 'before' rather than 'after', to match definition below?} 
each line of execution and log the variables' current values.

\noindent\textbf{Definition:} \textit{Program State.} Formally, we define the program state after the execution of a specific line, $l$, as $s(l)$, represented as a set of variable values at this moment:

\begin{equation*}
    s(l) = \{M(v, l)~|~v \in V, ~l \in L\}
\end{equation*}

$V$ represents the set of all traced variables, and $L$ is the set of lines with source code. Figure \ref{fig:program-states-trace} shows an illustrative example of a simple factorial program and the comments after the source code indicate the program state after the execution of that line. Also, we do not log the program state for lines without executable code, such as line-8 of Figure \ref{fig:program-states-trace}.

\begin{figure}[h]

  \centering
  \resizebox{0.9\linewidth}{!}{
  \begin{tabular}{l}
    \toprule
    %\underline{\small Source code w/ concrete trace}     \\
    {
\begin{lstlisting}[language=c-pretty]
// INPUT: 4
int factorial() { 
  int x, y; // {'x': 32767, 'y': 32767}
  x = atoi(argv[1]); // {'x': 4, 'y': 32767}
  if (x < 0) { // {'x': 4, 'y': 32767}
    y = -1;
    return y;
  }
  y = 1; // {'x': 4, 'y': 1}
  for (int i = 1; i <= x; i++) // {'x': 4, 'y': 24, 'i': 5}
  {
    y *= i; // {'x': 4, 'y': 24, 'i': 5}
  }
  return y; // {'x': 4, 'y': 24, 'i': 5}
}
\end{lstlisting}}
    \\
    
    \bottomrule
  \end{tabular}}
  \vspace{-1mm}
  \caption{\small Program states with concrete runtime values.}
  
\label{fig:program-states-trace}
\end{figure}

Note that a source code line could be executed multiple times due to a loop or recursion. While a more detailed representation of program execution might provide additional insights, it also increases the complexity and computational requirements of the model. As a trade-off between the complexity and performance, we use the last occurring execution of each line to finalize the program states, so that $s(l)$ keeps getting updated until the execution terminates. 

We apply such a trade-off based on the observations of real executions.
Specifically, the last occurring values are typically sufficient to capture the results of loops and recursions.
For example, when calling a recursive function, only the last occurring value(s) of returned variable(s) will be taken to fulfill the following execution of the caller. Similarly, the final values when loops finish will take part in the future execution.
As shown in line-12 of Figure \ref{fig:program-states-trace}, variable {\tt y} gets multiplied inside a loop to calculate the factorial. Its value changes in each iteration, but it is less informative to reason about the program's overall behavior, as only the final value is used as the return value (line-14). Thus, we would represent {\tt y} using the value from the last occurring execution of the loop.

\subsection{Quantized Variable Values}
\label{subsec:quantized_value_def}

As we introduced in \S\ref{sec:intro}, the distribution of concrete values is sparse and complex, consequently difficult for a statistical model to fit. In addition, concrete values are not always necessary. Some common program behaviors are accompanied by extremely large or small variable values -- for example, in C, uninitialized variables are often set to zero or uncommonly large variables, but the concrete values are not meaningful because they depend only on the data remaining on the stack, which could be randomly large or small. The model could represent such behaviors by estimating the value ranges of variables without accurately predicting their concrete values which are not informative or meaningful. Figure \ref{fig:program-states-trace} displays some of these cases: after the execution of line-3, {\tt x} and {\tt y} are uninitialized and randomly initiated as 32,767, which has no concrete meaning but only makes the training data noisy and sparse.

\begin{table}[h]
\centering
\caption{\small \tool's design of quantized variable values.}
\small
\resizebox{\linewidth}{!}{
\begin{tabular}{l|l|r|r}\hlineB{3}
\textbf{Data Type} & \textbf{Value Types} & \textbf{Concrete Value} & \textbf{Quantized Value}\\\hlineB{3}
\multirow{18}{*}{Basic} & \multirow{5}{*}{Integer} & $0 < v \leq 10,000$ & Positive Regular \\\cline{3-4}
 &  & $10,000 < v$ & Positive Large\\\cline{3-4}
 &  & 0 & Zero\\\cline{3-4}
 &  & $-10,000 \leq v < 0$ & Negative Regular \\\cline{3-4}
 &  & $v < -10,000$ & Negative Large\\\cline{2-4}
 & \multirow{7}{*}{Float/Double} & $0.0 < v \leq 1.0$ & Positive Small\\\cline{3-4}
 & & $1.0 < v \leq 10,000.0$ & Positive Regular\\\cline{3-4}
 & & $10,000.0 < v $ & Positive Large\\\cline{3-4}
 & & $0.0$& Zero\\\cline{3-4}
 & & $-1.0 < v < 0$ & Negative Small\\\cline{3-4}
 & & $-10,000.0 \leq v < -1.0$ & Negative Regular\\\cline{3-4}
 & & $ v < -10,000.0 $ & Negative Large\\\cline{2-4}
 & \multirow{3}{*}{Character} & `$\backslash$0' & Null\\\cline{3-4}
 &  & $v \in$ \{a-zA-Z\}& Alphabetic\\\cline{3-4}
 &  & $v \neq `\backslash 0`; v \notin$ \{a-zA-Z\}& Non-alphabetic\\\cline{2-4}
 & \multirow{2}{*}{Boolean} & 0& False\\\cline{3-4}
 &  & 1& True\\\cline{2-4}
 & Void & -& Void\\\hlineB{3}
\multirow{6}{*}{Array} & \multirow{2}{*}{Integer} & $[v_1, v_2, \ldots, v_n]$; & Initialized\\\cline{4}
 & & $quantize(v_i) \in \text{Integer}$ & Not Initialized\\\cline{2-4}
 & \multirow{2}{*}{Float/Double} & $[v_1, v_2, \ldots, v_n]$; & Initialized\\\cline{4}
 & & $quantize(v_i) \in \text{Float/Double}$ & Not Initialized\\\cline{2-4}
 & \multirow{2}{*}{Character} & \multirow{2}{*}{``$\langle$string$\rangle$''}&
 Initialized\\\cline{4}
 & & & Not Initialized\\\hlineB{3}
\multirow{6}{*}{Pointer} & \multirow{2}{*}{Integer} & 0x0& Null\\\cline{3-4}
 &  & Not 0x0& Not Null\\\cline{2-4}
 & \multirow{2}{*}{Float/Double} & 0x0& Null\\\cline{3-4}
 & & Not 0x0& Not Null\\\cline{2-4}
 & \multirow{2}{*}{Character} & 0x0& Null\\\cline{3-4}
 & & Not 0x0& Not Null\\
\hlineB{3}
\end{tabular}
}

\label{tab:def_quantized_value}
\end{table}

To reduce the data complexity and increase the density, we define 30 categories for quantized values in Table~\ref{tab:def_quantized_value}. To comprehensively represent the variable values, the proposed quantized categories consider both types, \ie the data types and value types, that are statically defined, and the dynamic runtime values. 
Our quantized categories cover the most common variable types and value types, 
which we have found sufficient to capture important program execution behaviors and relationships. By focusing on the most frequent value types, we can capture the essential features of program execution. This makes our approach effective at capturing the generalized program execution behaviors and patterns. We empirically illustrate our quantization strategy's effectiveness in \S~\ref{subsec: rq_quantized_value}.

\subsection{Building Learnable Labels for Code Models}
\label{subsec:build_labels}

We used supervised pre-training with traces. We construct labels for code models to learn two main perspectives of execution: program states and execution coverage.

\paragraph{\textbf{Program State Labels.}}
As we discussed in previous sections, we first trace the program variables during execution and log their runtime values. We then quantize these values into pre-defined categories. This process results in a sequence of program states, each represented by a set of quantized variable values (as shown in Figure~\ref{fig:program-states-trace}), and we build the learnable features for the code model on top of these program states. Specifically, we build labels for variables that can be quantized into Table~\ref{tab:def_quantized_value}'s categories and train the model to predict these labels given their source code representations (\S~\ref{subsubsec:learning_exec_rep}). The label for each variable is represented as a tuple: \emph{(data type, value type, quantized value)}. For example, in Figure~\ref{fig:program-states-trace}, the label of variable \texttt{X} occurring at line-3 is \emph{(Basic, Integer, Positive Large)}, as the current value of \texttt{X} is 32,767. We build such labels for all occurrences of valid variables that can be quantized, and the set combining all labels is considered as the program state labels of the code sample.

\paragraph{\textbf{Execution Coverage Labels.}} 
To unify our design and reduce the complexity of the model's learning process, we also build execution coverage labels for each occurrence of variables, aligning with the program state labels. Specifically, we represent the coverage of a variable with a binary label, ``Yes" or ``No". The variables within the executed line will be labeled as ``Yes", and those within the unexecuted lines will be labeled as "No" and further assigned with an ``Unknown" quantized value while retaining their labels for their data type and value type. For example, in Figure \ref{fig:program-states-trace}, Line-6 is not executed, so {\tt y} at this line has the coverage label of ``No" and the program state label of (Basic, Integer, Unknown), while {\tt y} at line-9 has the coverage label of ``Yes" and the program state label of (Basic, Integer, Positive Regular).

\section{Model}
In this section, we explain the details of \tool's components and learning objectives during pre-training and fine-tuning. 

\begin{figure}[th]
    \centering
    \includegraphics[width=\columnwidth]{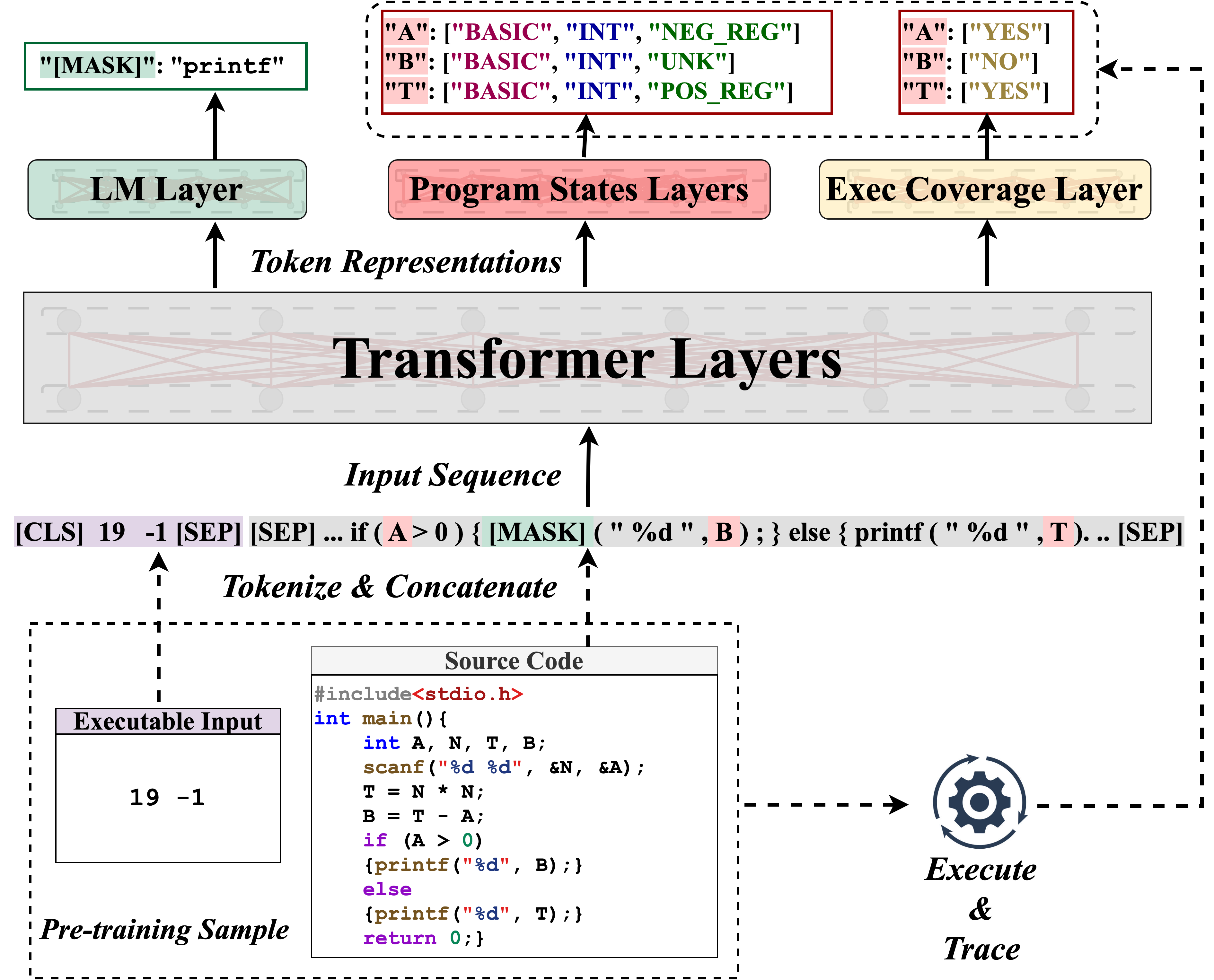}
    \caption{\small High-level model architecture of \tool. In the labels for program state layers, NEG\_REG means ``Negative Regular", UNK means ``Unknown", and POS\_REG means ``Positive Regular", which we have defined in Table~\ref{tab:def_quantized_value}.}
    \label{fig:model_arch}
    %\vspace{-5mm}
\end{figure}

\paragraph{\textbf{Model Architecture.}} Figure~\ref{fig:model_arch} shows the high-level architecture of \tool's pre-training. The backbone of \tool is a 12-layer Transformer encoder, similar to BERT~\cite{devlin-etal-2019-bert} and RoBERTa~\cite{liu2019roberta}, which learns the generic code representations. On top of the backbone Transformer layers, \tool stacks multiple multi-layer-perceptron (MLP) layers as prediction heads for different tasks. During the pre-training, as shown in Figure~\ref{fig:model_arch}, \tool applies a language model prediction head, \ie LM layer, to predict the masked token given its contextualized representation, a program state prediction head to predict the program states labels that we defined in \S~\ref{subsec:build_labels}, and an execution coverage head to prediction the execution coverage labels. For the task-specific fine-tuning, the backbone Transformer layers are loaded with the pre-trained weights, while the prediction heads are replaced by a newly initialized head customized for the specific downstream task.

\subsection{Execution-aware Pre-training}
\subsubsection{Model Input of Pre-training}
Each pre-training sample includes the source code of an executable program and a valid executable input. As shown in Figure~\ref{fig:model_arch}, the executable input and the source code are flattened and concatenated as one sequence. To distinguish the input from the source code, as they are different modalities, \tool uses special \texttt{[SEP]} tokens to separate them and indicate individual positions. To alleviate the out-of-vocabulary concern of programming languages~\cite{karampatsis2020bigcode}, \tool takes a pre-trained SentencePiece~\citep{kudo-richardson-2018-sentencepiece} subword tokenizer with vocabulary size of 50,000. It uses this tokenizer to divide the concatenated sequence into a new sequence of sub-tokens.

Formally, we define the executable inputs as $E = \{e_1, ..., e_i\}$ and flattened source code as $C = \{c_1, ..., c_j\}$, then the final model input will be $\mathcal{I} = $\{\texttt{[CLS]}$, e_1, ..., e_i, $\texttt{[SEP]}$, $\texttt{[SEP]}$, c_1, ..., c_j, $\texttt{[SEP]}\}. \tool truncates the executable inputs and the source code separately if they are too long. \tool sets the maximum length of the executable input sequence to 64 tokens, and the source code to 960 tokens. These numbers are selected based on the statistics of executable inputs' length of our pre-training dataset (\S\ref{subsubsec:pretrain_dataset}), and fit the rest of the model input with source code.

Note that the execution traces are not part of the model input, but are used as ground truth labels for the model to predict during pre-training.

\subsubsection{Learning Execution-aware Code Representations with Traces}
\label{subsubsec:learning_exec_rep}
\tool is pre-trained with multiple objectives to jointly capture the static and dynamic perspectives of the source code.

\paragraph{\textbf{Learning Code Text.}} Learning code text is the essential first step toward understanding the execution of a program, as code text is the primary source of capturing the code naturalness~\cite{hindle2012naturalness} and other static properties.
We implement the code text learning objective by adapting the masked language model objective~\cite{liu2019roberta, devlin-etal-2019-bert, feng2020codebert}. Specifically, given the model input sequence, $\mathcal{I}$, \tool randomly chooses 15\% of tokens~\cite{liu2019roberta, devlin-etal-2019-bert} only from the source code sequence $C$ part and replaces with the special \texttt{[MASK]} token (\eg \texttt{printf} in Figure~\ref{fig:model_arch} is masked). It leaves the executable input sequence $E$ as is. The model is trained to encode the context of \texttt{[MASK]} into its code representation, $r_{masked}$, and reconstruct the concrete masked tokens conditioned on the representation. We represent the loss of learning code text as:

\begin{equation}
\label{eq:mlm_loss}
    \mathcal{L}_{code-text} = \sum_{masked} - log P(c_{masked} ~| ~r_{masked})
\end{equation}

In Figure~\ref{fig:model_arch}, the LM (Language Model) layer receives the masked token representation generated by the last Transformer layer. The LM layer then predicts the concrete tokens by mapping the token representation to the probability of each token in the vocabulary, using an MLP (Multi-Layer Perceptron) layer. This process can be thought of as a classification task, where the number of classes is equal to the size of the vocabulary. The goal is to learn a mapping from the masked token representation to the most probable token in the vocabulary, given its context.

\paragraph{\textbf{Learning Program States.}} The second pre-training objective, program state prediction (PSP), is designed to enable the model to learn program execution behavior by predicting the program state labels of the traced variables. These program state labels, as defined in \S\ref{subsec:build_labels}, contain information about the data types, value types, and quantized values of the variables at the end of the program execution. Specifically, \tool first identifies the variable tokens in the source code sequence, denoted as $\{c_{var}~|~c_{var}\in V\} \subseteq C$, where $V$ is the set of all traced variables and $C$ is the source code sequence. It then extracts the representation, $r_{var}$, of each variable token and feeds it into the program state layer. The program state layer predicts the variable's joint likelihood of being the ground-truth data type, $d_{var}$, value type, $t_{var}$, and quantized value, $q_{var}$. Note that if a variable is tokenized as multiple sub-tokens, all belonging sub-tokens share the same program state label. Finally, \tool computes the loss of PSP as the sum of the losses of all variable tokens used for predicting their program states. Mathematically, the loss is expressed as follows:

\begin{equation}
\label{eq:program_state_loss}
    \mathcal{L}_{program-state} = \sum_{var} - log P(~d_{var}, t_{var}, q_{var} ~| ~r_{var})
\end{equation}

\paragraph{\textbf{Learning Variable Coverage.}} The third pre-training objective, variable coverage prediction (VCP), aims to learn the execution coverage, which is crucial for understanding the control flow of the code given a specific input. Similar to the PSP objective, VCP targets making predictions for variable tokens. The label of variable coverage is binary, where 1 represents the line this variable belongs to is covered, otherwise 0. Also, sub-tokens belonging to the same variable will be assigned the same coverage label. The loss of VCP is as follows:

\begin{equation}
\label{eq:cov_loss}
    \mathcal{L}_{var-cov} = \sum_{var} - log P(cov_{var} ~| ~r_{var})
\end{equation}

Finally, \tool combines the losses of all three objectives and computes their sum as the final loss of a pre-training sample. It back-propagates the gradients through both the prediction layers and the backbone Transformer layers to update their weights. We denote the full set of \tool's learnable parameters as $\theta$ and represent the loss as follows:

\begin{equation}
    \mathcal{L}(\theta) = \mathcal{L}_{code-text}(\theta) + \mathcal{L}_{program-state}(\theta) + \mathcal{L}_{var-cov}(\theta)
\end{equation}

\subsection{Task-specific Fine-tuning}
\tool loads the model weights of Transformers layers, which are pre-trained to produce execution-aware code representations, and further fine-tunes the model for downstream tasks. We consider three downstream tasks as the main applications for \tool: (1) Static estimation of program execution which includes both execution coverage prediction and runtime variable value prediction; (2) Semantic Clone Retrieval; (3) Vulnerability Detection.

\paragraph{\textbf{Static Execution Estimation.}} Our goal of pre-training is to encode the execution patterns into the code representation, so the model could estimate the program execution statically. As a direct application, \tool fine-tunes the model to predict (1) the execution coverage and (2) runtime variable values using source code and program input. \tool evaluates the fine-tuned model to estimate the execution of unseen programs in the same way.

Specifically, for execution coverage prediction, \tool identifies all the branching statements to locate the branches, $B = \{b_1, b_2, ..., b_m\}$, within the source code. It trains the model to predict a binary label, 0 means the branch is not covered by the current execution and 1 means covered, for each $b_i \in B$. For the model's convenience to make predictions, the special token \texttt{[MASK]} is inserted at the beginning of each branch. For example, the following \texttt{if-else} has two branches that are pre-processed for branch prediction: \texttt{if (condition) \{[MASK] ...\} else \{[MASK] ... \}}. During the fine-tuning, the Transformer layers learn to encode the branch information into the corresponding \texttt{[MASK]} token representation with the built-in bi-directional attention and positional encoding. Then the classification head takes \texttt{[MASK]} representations to predict whether a branch is covered by the current execution. For variable value prediction, \tool identifies variables, $V = \{v_1, v_2, ..., v_n\}$ and trains the model to predict their quantized values (\S\ref{subsec:quantized_value_def}) during the execution.

\paragraph{\textbf{Semantic Clone Retrieval.}} Detecting semantic clones is significant for software maintenance~\cite{kim2017vuddy,li2016vulpecker}, yet very challenging in practice since the token and syntactic structures overlap among semantic clones may be quite limited. This task requires the model to estimate the program behaviors without executing the programs and capture the similarity among them. It evaluates the model's semantic reasoning capacity to identify the code similarity and retrieve clones: given a program as a query, and an arbitrary collection of programs as candidates, the model needs to identify the query's semantic clones from possibly thousands of candidates.

\paragraph{\textbf{Vulnerability Detection.}} Vulnerability detection is a crucial task in software security, aiming to identify potential security vulnerabilities in software code that could be exploited by attackers. The vulnerabilities may exist due to various reasons, including programming errors, design flaws, or configuration issues. Detecting these vulnerabilities early in the software development lifecycle can prevent potential attacks, mitigate risks, and save resources.  We fine-tune \tool's pre-trained model on datasets consisting of vulnerable and non-vulnerable code samples, so the model learns to classify code functions as vulnerable or non-vulnerable by estimating their execution behavior.

\section{Experimental Setup}

\subsection{Trace Collection}
In this section, we explain how we traced the dynamic information in programs to produce concrete traces, given the source code and program input.

First, we compile the program using \texttt{gcc} with
the options \texttt{-g -O0}. Option \text{-g} preserves debug information, which is necessary in order to read variables and source code locations using the debugger, and option \texttt{-O0} disables compiler optimizations, which could optimize out some variables thus preventing them from being read at runtime. We use this option because we seek to model the semantics of the \textit{source code} in terms of variable values rather than the optimized machine code.

Second, we load the program with the given standard input redirected to \texttt{stdin} and attach the \texttt{gdb}\footnote{\url{https://www.sourceware.org/gdb}} debugger, using the Python API to implement the tracing command.
Starting from the entry point (\texttt{main}), we execute the program one line at a time using the \texttt{step} command.
At each line, we print out the concrete values of all variables in scope.
We also set breakpoints at the entry of each user-defined function, where we log the values of each parameter.
For numeric types, we simply log their string representation. For \texttt{char} and \texttt{char *} (string) types, we log the human-readable values of the chars/strings. We use gdb's pretty-printer to print \texttt{struct} types and statically allocated array types, such as \texttt{int[<size>]}. For pointer types, we print the memory address of the pointer as a hex code.  We only traced the functions that were defined in the source code and skipped over all standard library functions.

\subsection{Dataset}

\subsubsection{Pre-training Dataset}
\label{subsubsec:pretrain_dataset}
IBM's CodeNet Dataset \cite{puri2021codenet} includes 4,053 programming challenges for several programming languages from the AIZU Online Judge and AtCoder platforms, and each problem has up to thousands of implementations submitted by distinct programmers. In this work, we focus on the C language as the main resource for the pre-training and downstream tasks, so we build our pre-training dataset with programming challenges that have C solutions. Besides the large number of samples and the complexity of programming challenges, we choose CodeNet to build our datasets as it maintains at least one and at most twenty executable inputs for each challenge, so we could execute and trace the implementations of the challenge, and consequently build our execution labels for the model to learn.

Out of 1,900 programming challenges with C solutions, we select 1,805 of them to build the pre-training dataset and leave the other 95 problems as held-out problems for evaluating the model's capacity for the downstream static execution estimation task. Splitting samples strictly by challenge effectively avoids the issue of data leakage from the training set to the held-out set.  We randomly sample up to 200 execution traces for each challenge, and this ends up with 121,319 training traces.

\subsubsection{Downstream tasks}
\label{subsubsec:downstream_task_def}
In this section, we introduce the datasets we use for each downstream task and explain the corresponding evaluation metrics. The statistics of these datasets are in Table~\ref{tab: downstream_dataset}.

\vspace{-1mm}
\paragraph{\textbf{Static Execution Estimation.}} We build the dataset for this task using CodeNet. We build the training samples from the 1,805 challenges that have been selected by the pre-training, and build evaluation samples from the held-out 95 challenges to avoid model memorization and data leakage. 

\noindent\textbf{Metrics.} For the execution coverage prediction, we consider evaluation metrics in two granularities: full execution path and branch coverage. Concretely, for a sample with $m$ branches, we denote the full set of their labels as $LB = \{lb_1, lb_2, ..., lb_m\}$, and the model prediction set as $\hat{LB} = \{\hat{lb_1}, \hat{lb_2}, ..., \hat{lb_m}\}$. If $LB == \hat{LB}$, we regard the prediction as matching the full execution path. For the branch coverage, we compute the occurrence of $lb_i == \hat{lb_i}$, where $1 \leq i \leq m$, and report the accuracy, precision, recall, and F1. Similarly, for the $n$ quantized variable values within the program, $QV = \{qv_1, qv_2, ..., qv_m\}$, our model makes predictions as $\hat{QV} = \{\hat{qv_1}, \hat{qv_2}, ..., \hat{qv_m}\}$. If $QV == \hat{QV}$, we say the model accurately predicts the full execution. For the individual value match, we compute the occurrence of $qv_i == \hat{qv_i}$ and report the accuracy.

\vspace{-1mm}
\paragraph{\textbf{Semantic Clone Retrieval.}} We use CodeXGLUE-POJ104~\cite{mou2016convolutional, msr2021codexglue} as the dataset for this task. CodeXGLUE-POJ104 contains 104 programming challenges, and each has 500 C/C++ solutions submitted by different programmers. CodeXGLUE~\cite{msr2021codexglue} reconstructs it as a public benchmark by splitting the dataset into Train (64 challenges), Dev (16 challenges), and Test (24 challenges) sets, with no overlapped challenge between any two sets. 

\noindent\textbf{Metrics.} MAP@R (Mean Average Precision @ R)\footnote{\url{https://en.wikipedia.org/wiki/Evaluation_measures_(information_retrieval)\#Mean_average_precision}} is the main metric of this task, where we follow the design of the CodeXGLUE benchmark. Average precision at R is a common metric to evaluate the quality of information retrieval; it measures the average precision scores of a set of the top-R clone candidates presented in response to a query program. The "R" for CodeXGLUE is 499 as it has 500 solutions for each challenge.

\vspace{-1mm}
\paragraph{\textbf{Vulnerability Detection.}} We utilized three publicly available datasets: REVEAL (RV)~\cite{chakraborty2021reveal}, D2A~\cite{zheng2021d2a}, and CodeXGLUE-Devign (CXG)~\cite{msr2021codexglue, Zhou2019DevignEV}. The REVEAL dataset was curated by Chakraborty \etal to simulate a real-world scenario where bugs are relatively rare, resulting in a ratio of approximately 1:10 between buggy and benign samples. The D2A dataset is a balanced dataset focusing on bug-fixing commits. It labels the previous version of modified functions as buggy and the fixed version as benign. Finally, the CodeXGLUE-Devign dataset, introduced by Zhou \etal, is also a balanced dataset that has been reconstructed as a public benchmark by CodeXGLUE, ensuring that all models can be evaluated using the same train/valid/test splits. 

\noindent\textbf{Metrics.} REVEAL is an imbalanced dataset, so we use F1 as the evaluation metric. D2A and Devign are balanced datasets, so we follow the original benchmark to report the classification accuracy.

%\vspace{-2mm}
\begin{table}[h]
\centering
\caption{\small Details of downstream tasks datasets.}\label{tab: downstream_dataset}
\small
\vspace{-2mm}
\resizebox{\linewidth}{!}
{
\begin{tabular}{l|c|r|r|r}\hline
Task &Dataset &Train &Valid &Test \\\hline
Execution Estimation & CodeNet & 121,319 & 13,116 & 13,116\\\hline
Clone Detection &CXG-POJ104 &32,000 &8,000 &12,000 \\\hline
\multirow{3}{*}{Vulnerability Detection} &REVEAL &15,867 &2,268 &4,535 \\
&D2A &4,644 &597 &619 \\
&CXG-Devign &21,854 &2,732 &2,732 \\\hline
\end{tabular}}
\end{table}

\subsection{Model Configuration}
\tool's backbone is a standard RoBERTa\textsubscript{BASE} architecture~\cite{liu2019roberta} with 12 layers of Transformer-encoder, and each layer has 12 attention heads and the hidden dimension is 768. \tool is initialized with the pre-trained weights from UnixCoder~\cite{guo2022unixcoder}\footnote{Specifically, we load unixcoder-base-nine, as its pre-training considers C language code samples: \url{https://huggingface.co/microsoft/unixcoder-base-nine}. Note that this checkpoint is pre-trained only with the MLM objective, while the original paper~\cite{guo2022unixcoder} reports other better-performing variants that are not released publicly.}, and we use its BPE tokenizer to split the rare tokens into BPE sub-tokens. The maximum sequence length is 1024 BPE tokens, and the longer sequence will be truncated. When the code sample is paired with executable inputs, the maximum length for the executable input is 64, and the source code is 960. Our experiments are conducted on 2 $\times$ 24GB NVIDIA GeForce RTX-3090 GPUs. We further pre-train the model for 10 epochs to learn the program execution with two learning rates, 5e-5 and 2e-5, and report the best-performing models for downstream tasks. For all the fine-tuning tasks, we use the learning rate of 8e-6. Learning rates typically decrease for later phases~\cite{feng2020codebert, guo2021graphcodebert, ding2021disco}, so \tool follows the same design. We use Adam optimizer~\cite{Kingma2015AdamAM} with the linear learning rate decay. Our model is implemented mainly with Pytorch~\cite{paszke2019pytorch} and Huggingface~\cite{wolf-etal-2020-huggingface}.
\section{Evaluation}
\label{sec:evaluation}

In this section, we ask the following four RQs:
\begin{itemize}[leftmargin=*]
    \item {\textbf{RQ1:}} How effective is \tool in statically estimating the program execution? 
     \item {\textbf{RQ2:}} How does our proposed training strategy contribute to learning the program execution?
     \item {\textbf{RQ3:}} Is our proposed quantized values for programs effective in guiding the model to learn program executions?
    \item {\textbf{RQ4:}} How does \tool perform \wrt statically pre-trained baselines for code understanding tasks?
\end{itemize}

\subsection{RQ1. Effectiveness of \tool in Static Estimation of Execution}
In this section, we demonstrate the effectiveness of \tool in statically estimating program execution. 
The evaluation is more challenging and realistic than \tool's pre-training as it requires the model to predict not only for individual variables but also branches and the full execution path.

\noindent\textbf{Baseline.} In this RQ, we mainly compare the execution-aware \tool with UnixCoder~\cite{guo2022unixcoder}. Now we explain the reasons for this choice. First, \tool is initialized with the pre-trained UnixCoder weights, so comparing \tool with the UnixCoder performance is a direct assessment of the impact of our proposed pre-training. Second, UnixCoder reports the state-of-the-art performance in many tasks, including clone detection, code search and summarization, and code generation and completion, significantly outperforming other pre-trained code models, such as CodeBERT~\cite{feng2020codebert} and GraphCodeBERT~\cite{guo2021graphcodebert}. Third, it consumes up to 1,024 tokens, while most pre-trained code models~\cite{feng2020codebert, buratti2020cbert, guo2021graphcodebert, ahmad2021plbart, wang2021codet5} take at maximum 512 tokens. By consuming longer sequences, UnixCoder is able to handle longer programs and make complete predictions without truncating code in many cases. As \tool is also designed to consume 1,024 tokens, it is not fair to compare it in this task with baselines with a maximum length of 512, as the baselines will necessarily consider fewer branches for prediction.

\vspace{-1mm}
\begin{table}[!htp]\centering
\caption{\small Performance on static execution estimation.}\label{tab: comp_branch_value}
\small
\resizebox{\linewidth}{!}
{
\begin{tabular}{l|c|c|c|c|c|c|c}\hlineB{2}
\multirow{3}{*}{\textbf{Model}} & \multicolumn{5}{c|}{\textbf{Coverage}} & \multicolumn{2}{c}{\textbf{Runtime Value}} \\\cline{2-8} & \textbf{Full Path}  &\multicolumn{4}{c|}{\textbf{Branch}} & \textbf{Full Exec}  &\textbf{Var} \\\cline{2-8}
&\textbf{Acc} &\textbf{Acc}&\textbf{Prec}&\textbf{Rec} &\textbf{F1} &\textbf{Acc} &\textbf{Acc}
\\\hlineB{2}
UnixCoder &63.7 &79.7 &81.7 &85.4&83.5 & 39.3 & 87.8\\\cline{1-8}

\tool &\textbf{71.6} &\textbf{83.1} &84.6 &\textbf{88.1} &\textbf{86.3}& \textbf{49.2} & \textbf{89.2} \\

-w/o MLM &70.4 &82.6 &\textbf{85.3} &86.0 &85.6& 49.0 & \textbf{89.2} \\
-w/o PSP &69.0 &81.4 &83.0 &86.9 &84.9& 44.0 & 87.4 \\
-w/o VCP &66.1 &80.3 &82.4 &85.6 &84.0& 46.7 & 89.0\\
-MLM-only &65.6 &81.0 &83.1 &86.0 &84.6& 43.0 & 87.5 \\
\hlineB{2}
\end{tabular}
}
\vspace{-2mm}
\end{table}

\noindent\textbf{Result.} The comparison is shown in Table~\ref{tab: comp_branch_value}, Row-1 \vs Row-2. \tool significantly outperforms UnixCoder in the static estimation of execution coverage and dynamic values of variables, especially when the evaluation granularity is coarse, \ie full execution path (Full Path column in Table~\ref{tab: comp_branch_value}) and the runtime values of the full execution (Full Exec column in Table~\ref{tab: comp_branch_value}). \tool correctly predicts the complete execution paths for 71.6\% held-out samples and accurately predicts all variable values for 49.2\% executions, revealing the execution-aware pre-training improves over UnixCoder's performance by 12.4\% and 25.2\%, respectively.

\noindent\textbf{Case Study with Qualitative Examples.} We present two qualitative examples in Figure~\ref{fig:qual_ex_cov} and \ref{fig:qual_ex_value} to concretely compare \tool with UnixCoder in execution coverage and runtime value predictions, respectively. Both samples have simple execution logic from the human perspective, but the statically pre-trained UnixCoder still fails to correctly estimate them. Figure~\ref{fig:qual_ex_cov} illustrates that UnixCoder is not sensitive to distinct inputs that trigger different execution coverage, while \tool is able to determine the numerical relations among varied values. Figure~\ref{fig:qual_ex_value} illustrates \tool's capacity in exposing abnormal program behaviors.

\begin{figure}[h]
  \centering
  %\resizebox{0.85\linewidth}{!}{
  \begin{tabular}{p{0.9\linewidth}}
    \toprule
    %\underline{\small Source code w/ concrete trace}
    \vspace{-4mm}
    {
\begin{lstlisting}[language=c-pretty, numbers=none]
//Input: 19 100
#include <stdio.h>
int main(){
    int A, N, T, B;
    scanf("%d %d", &N, &A);
    T = N * N;
    B = T - A;
    if (A > 0) {printf("%d", B);} // Branch-1
    else {printf("%d", T);} // Branch-2
    return 0;
}
\end{lstlisting}}
    \\
    \underline{\small {\color{red}\textbf{UnixCoder Predictions (Wrong)}}}     \\
    {\small Branch-1: { Not executed}}\\
    {\small Branch-2: { Not executed}}\\
    \underline{\small { \textbf{\dkgreen {TRACED Predictions (Correct)}}}}     \\
    {\small Branch-1: { Executed}}\\
    {\small Branch-2: { Not Executed}}\\
    %{\lstinputlisting[language=c-pretty]{figures/quantized_program_state.c}}
    
    \bottomrule
  \end{tabular}
  \caption{\small The qualitative example of execution coverage prediction. The source code is the same as Figure~\ref{fig:motivation}, but the input triggers a different execution path. \tool correctly flips the prediction while UnixCoder remains the same prediction.}
  \label{fig:qual_ex_cov}
\end{figure}

\begin{figure}[h]
  \centering
  %\resizebox{0.85\linewidth}{!}{
  \begin{tabular}{p{0.9\linewidth}}
    \toprule
    %\underline{\small Source code w/ concrete trace}
    \vspace{-4mm}
    {
\begin{lstlisting}[language=c-pretty, numbers=none]
//Input: 4 4320 4320 4320
#include <stdio.h>
int main (void) {
    int n, a, max = 0, sum = 0, i;
    for (i = 0; i < n; i++){ // Quantized value of n? 
        scanf("\%d", &a);
        if (a > max) max = a;
        sum += a;
    }
    printf("\%d\\n" , sum - max / 2);
    return 0;
}
\end{lstlisting}}
    \\
    \underline{\small {\color{red}\textbf{UnixCoder Prediction (Wrong)}}}     \\
    {\small n: Zero}\\
    \underline{\small { \textbf{\dkgreen {TRACED Prediction (Correct)}}}}     \\
    {\small n: Negative Large}\\
    %{\lstinputlisting[language=c-pretty]{figures/quantized_program_state.c}}
    
    \bottomrule
  \end{tabular}
  \caption{\small The qualitative example of runtime value prediction. The sample contains a vulnerability of type CWE-457 ``Use of Uninitialized Variable". The uninitialized \texttt{n}, which is randomly assigned as -32767, is used in the \texttt{for-loop}. \tool successfully exposes this abnormal behavior statically by identifying $n$ as a ``Negative Large" value while UnixCoder fails. Predictions of other variables are hidden for better illustration.}
  \label{fig:qual_ex_value}
\end{figure}

\begin{finding}
With a similar number of learnable parameters, \tool outperforms the state-of-the-art pre-trained code model in the static estimation of program execution task. Our proposed pre-training successfully encodes the execution awareness into \tool's code representations.
\end{finding}

\subsection{RQ2. Effectiveness of \tool's Pre-training Objectives}
\label{subsec: rq_training_objectives}
One of the main contributions of this paper is proposing multi-task pre-training to effectively learn the execution-aware code representations. In this RQ, we study the effectiveness and contribution of each of \tool's objectives, and consequently illustrate the importance of the multiple tasks. 

To conduct these experiments, we remove one pre-training objective at a time and pre-train the variant with exactly the same setup as the main model. Then we fine-tune the variant on the static execution estimation task and compare the performance with the main model. We also consider a variant that is pre-trained on our dataset but only with MLM objectives. The results are shown in Row 3-6 of Table~\ref{tab: comp_branch_value}. Removing any objective hurts \tool's performance, suggesting that comprehensively learning both static and dynamic code properties is more effective than learning one perspective alone.

\begin{finding}
\tool's multi-task pre-training helps the model comprehensively learn both static and dynamic aspects of source code. Removing any one of \tool's three pre-training objectives noticeably hurts the model's performance in statically estimating program executions.
\end{finding}

\subsection{RQ3. Effectiveness of \tool's Quantized Variable Values}
\label{subsec: rq_quantized_value}
Another contribution of this paper is that the simplified and compact representation of program executions helps code models to capture dynamic code properties. In this RQ, we empirically reveal that the design of quantized variable values especially contributes to the effective learning of the code models, as it reduces the data sparsity of variable values but still defines sufficiently detailed value categories to distinguish dissimilar values. 

To isolate the evaluation of \tool's quantized values, we pre-train several variants by only recreating quantized value labels, \ie $q_{var}$ in Equation~\ref{eq:program_state_loss}, using different value abstraction strategies. For example, when we pre-train a variant studying the impact of concrete values, we replace \tool's defined $q_{var}$ with the concrete traced values. As different strategies abstract values at different granularities, it is not feasible to compare them for the value prediction task, since the coarse-grained strategy will benefit. Therefore, we only fine-tune the studied variants for the execution coverage prediction.

\noindent\textbf{Baseline.} First, we consider comparing with concrete values, as it is the most intuitive strategy to represent variable values. Then, we consider two data abstractions from LExecutor~\cite{souza_lexecutor_2023}: coarse and fine-grained. They share similar high-level intuition with us, mapping concrete values to pre-defined bins to reduce data complexity and consequently help the model's learning. Note that LExecutor's data abstraction serves a different goal than \tool, and focuses on Python while \tool focuses on C, so we could not directly reuse their pre-defined bins. As their definition of data abstraction is clear and straightforward, we re-implement their data abstraction for the C language and integrate it into our framework for comparison. We discuss and compare LExecutor with \tool in more detail in the Related Work section (\S\ref{sec:related_work}). 

\begin{figure}[h]
    \centering
    \includegraphics[width=0.8\columnwidth]{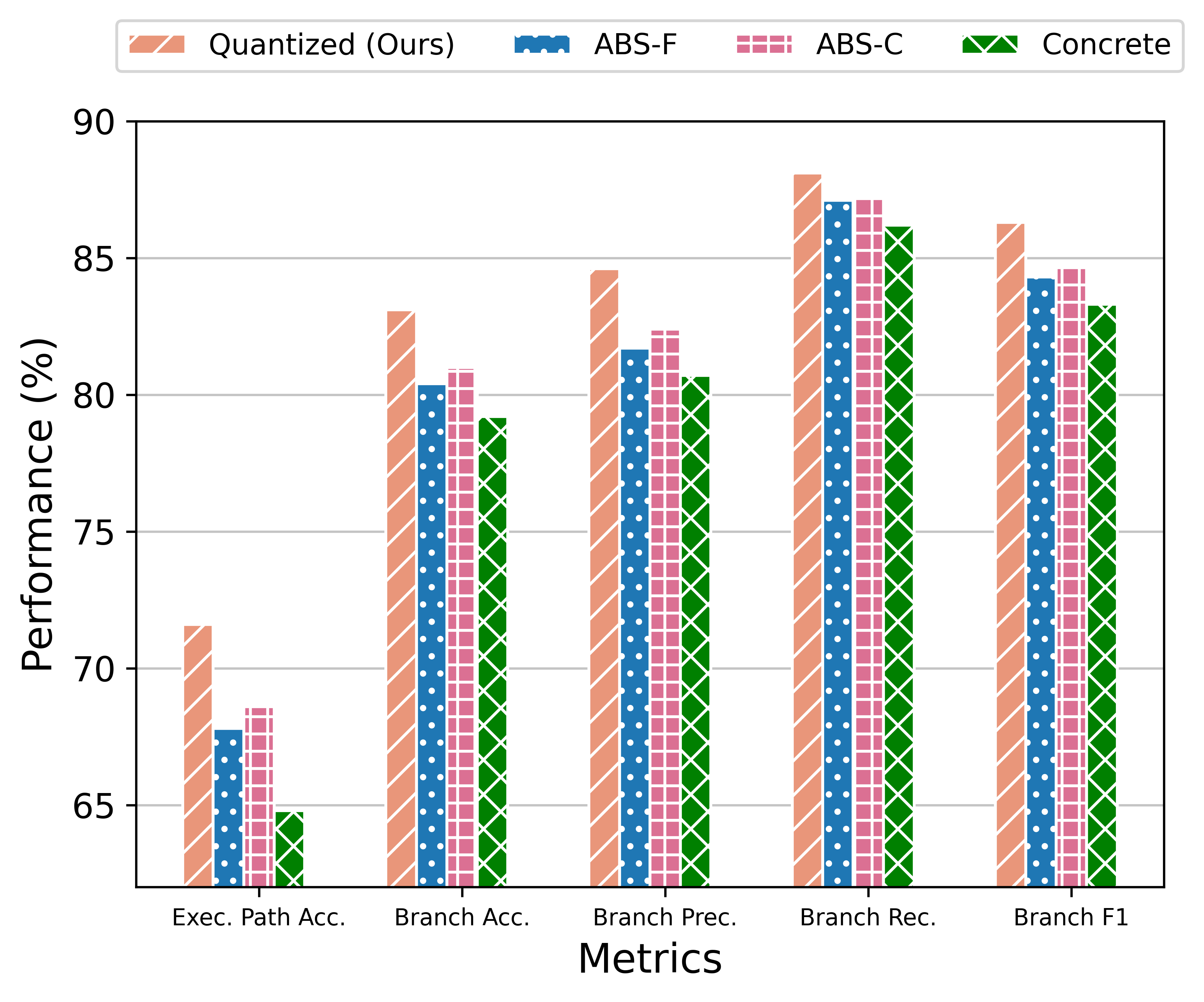}
    \caption{\small Comparing \tool's design of quantized variable values with other value abstraction strategies.}
    \label{fig:value_gran_hist}
\end{figure}

\noindent\textbf{Results.} The comparison of value abstractions are shown in Figure~\ref{fig:value_gran_hist}. Unsurprisingly, concrete values report poor performance compared to other data abstractions, empirically revealing the difficulties for code models to fit sparse and complex data distributions. Interestingly, we notice both of LExecutor's abstractions perform slightly worse than \tool. We speculate that LExecutor is not as sensitive as \tool to numeric relations in the conditional statements, as they do not distinguish among small, regular, and large values. Note that execution coverage is not the main focus of LExecutor, so more fine-grained categories are not required to serve its goal, while they are empirically proven to be necessary for \tool's scope.

\begin{finding}
\tool's quantized variable values directly contribute to the effectiveness of its execution-aware pre-training. It reduces the data sparsity of concrete values but defines sufficiently detailed value categories to distinguish dissimilar values for reasoning about execution paths.
\end{finding}

\subsection{RQ4. \tool's Performance in Code Understanding Tasks}
In this RQ, we study \tool's performance on two code understanding tasks: semantic clone retrieval and function-level vulnerability detection. Note that samples for these tasks are not paired with executable inputs, so the model needs to reason about the general code semantics to make predictions. 

\noindent\textbf{Baselines.} We consider five pre-trained code models with similar parameter sizes to \tool. CodeBERT~\cite{feng2020codebert} pre-trains a RoBERTa model with MLM and replaced token detection (RTD) tasks. GraphCodeBERT~\cite{guo2021graphcodebert} is initialized with CodeBERT and continues pre-training with augmented data flow graphs to learn the static data dependencies. PLBART~\cite{ahmad2021plbart} and CodeT5~\cite{wang2021codet5} both apply the seq2seq neural architecture, where PLBART adapts the BART~\cite{lewis-etal-2020-bart} model to learn code translation and summarization, and CodeT5 adapts~\cite{2020t5} to predict the missing code tokens and locate the identifiers. We also, again, consider UnixCoder as a baseline.

\begin{table}[!htp]\centering
\caption{\small Comparison of Clone Retrieval and bug detection.}\label{tab: rq_code_understand_task}
\small
\resizebox{\linewidth}{!}
{
\begin{tabular}{l|c|c|c|c}
\hlineB{2}
\textbf{Task} &\textbf{Clone Retrieval} &\multicolumn{3}{c}{\textbf{Vulnerability Detection}} \\\hline
\textbf{Dataset} &\textbf{POJ-104}  &\textbf{RV} &\textbf{D2A} &\textbf{CXG} \\
\hlineB{2}
\textbf{Metric} &\textbf{MAP@R}&\textbf{F1} &\textbf{Acc} &\textbf{Acc} \\
\hlineB{2}
CodeBERT & 82.7
& 47.3& 59.2& 63.4\\
GraphCodeBERT & 86.7
& 46.6& 61.0& 62.9\\
PLBART-base & 75.9
& 46.9& 61.7 & 63.3\\
CodeT5-base* & 65.9
& 46.5& \textbf{62.1}& 64.4\\
UnixCoder & 89.5
& 47.4& 61.2& 65.3\\\hlineB{2}
\rowcolor{rowgray}
\tool & \textbf{91.2}& 
\textbf{50.4}& \textbf{62.1}& \textbf{65.9}\\
\hlineB{2}
\end{tabular}}
\begin{flushleft}
\small
*CodeT5-base has 223M parameters, roughly twice as large as other baselines and \tool. We report its performance as CodeT5-small has only 60M parameters and performs poorly, and CodeT5 does not provide a $\sim$110M model.
\end{flushleft}

\end{table}

\noindent\textbf{Results.} We show the results in Table~\ref{tab: rq_code_understand_task}. Even though the samples in these benchmarks do not have executable inputs, \tool still outperforms the statically pre-trained models by a clear margin. We speculate the reason is that \tool could estimate the general execution behaviors without specific inputs, and the program semantics regarding these two code understanding tasks could be better captured with such a general sense. Specifically, clone retrieval requires the model to identify the behavioral similarities of code as semantic clones mostly differ in code text and syntax. Also, vulnerable code with potential anomalies could be directly identified by \tool in some cases like Figure~\ref{fig:qual_ex_value}.

\begin{finding}
\tool outperforms statically pre-trained models in clone retrieval and vulnerability detection tasks, suggesting \tool's general estimation of execution helps it capture the code semantics more effectively.
\end{finding}

\section{Related Work}
\label{sec:related_work}

\paragraph{\textbf{Pre-trained Models for Source Code}} The research community has shown a growing interest in developing pre-trained Transformer models for source code. These models can be broadly categorized into three primary architectures: Encoder-only~\cite{feng2020codebert, guo2021graphcodebert, xin2021syncobert,bui2021corder,cubert,buratti2020cbert, ding2021disco}, Decoder-only~\cite{chen2021codex,xu2022systematic, austin2021synthesis}, and Encoder-decoder~\cite{niu2022sptcode,ahmad2021plbart,guo2022unixcoder,Li2022CompetitionLevelCG, chakraborty2021natgen}. Encoder-only models predominantly employ MLM objective and sequence understanding tasks (\eg predicting next statement~\cite{cubert} and contrasting semantics~\cite{ding2021disco}). This architecture excels at understanding the static code features. Decoder-only models, on the other hand, are typically trained by predicting code tokens in a left-to-right manner. This architecture focuses on generating code text based on learned patterns. The Encoder-decoder models combine the strengths of both Encoder-only and Decoder-only models and are pre-trained using various tasks, including denoising autoencoding for reconstructing wrongly permuted tokens~\cite{ahmad2021plbart}, predicting missing identifiers in the code~\cite{wang2021codet5}, and recovering method names from the source code~\cite{niu2022sptcode}.

These models primarily focus on learning the static aspects of source code but often miss out on capturing the dynamic properties of code execution. This limitation restricts these models from accurately inferring runtime behaviors, debugging issues, and understanding complex program states.

\paragraph{\textbf{Modeling Program Execution}}

Pei et al. \cite{pei2020trex,pei_stateformer_2021,pei_neudep_2022} proposed a series of pioneering works to learn the executions of \emph{binary} programs with Transformer-based models.
They used concrete values from registers, which are feasible in their scope because binary programs have a smaller space of possible values and effects compared to source code.
On the other hand, our work focuses on encoding execution at the source code level by imitating the developers' code practice. Variables in source code have more complicated data and value types than machine registers. We introduce quantized values in order to decrease the data complexity and sparsity.

Several works \cite{zaremba_learning_2015,reed_neural_2016,bieber_learning_2020,nye_show_2021,bieber_static_2022} have attempted learning to execute programs as a direct goal. Souza and Pradel \cite{souza_lexecutor_2023} also proposed LExecutor to predict missing values during execution. While it shares similar intuition of mapping concrete values to discrete categories, LExecutor is distinct from \tool in several perspectives. First, LExecutor focuses only on predicting the values, while \tool proposes a general pre-training strategy to encode the comprehensive execution awareness, not only values but also execution coverage, into the code representation. Besides, to yield code representations at a better quality, \tool jointly learns both code text and dynamic executions rather than sticking to a single perspective. Due to the distinct aims and designs, we empirically illustrate in RQ3 (\S\ref{subsec: rq_quantized_value}) that LExecutor's value abstractions are not perfectly aligned with our scope.

Nie et al. \cite{nie_learning_2023} annotated programs with information about the program's possible executions without executing the code but provided only statically available information. Conversely, several works \cite{henkel_code_2018,wang_dynamic_2018,wang_blended_2020,patra_nalin_2022} require dynamic traces as input.
We show that \tool's pre-training is able to encode the execution awareness into code representation and estimate the dynamic semantics with static information alone.

\section{Threats to Validity}
\label{sec:threats}

\paragraph{Internal Validity} First, the current design of quantized value is not covering all variables within the program due to the complexity of their data structures, value ranges, and/or memory allocations. Second, currently, we only trace the program by feeding it valid and executable inputs which will not terminate the program or throw errors. This might make the model less capable of capturing program termination and error-throwing behaviors.

\vspace{-2mm}
\paragraph{External Validity} At present, \tool supports only the C programming language. This limitation is due to the reliance on the capabilities of the tracer used to log the execution history, which may not be readily available or equally effective for other programming languages. In order to extend \tool's applicability, it is necessary to ensure that the tracer employed can accurately and consistently capture the required information across different languages. Adapting \tool to multiple languages would require the development or adaptation of tracers that can effectively handle the intricacies of each language and produce comparable results, enabling a consistent analysis of code behavior across a broader range of programming languages. 

\section{Conclusion} In this paper, we propose \tool, an execution-aware pre-trained model that jointly learns the static and dynamic code properties, to address the limitation of existing, statically pre-trained code models. The evaluation empirically reveals that \tool is more effective in estimating code execution statically than statically pre-trained models. \tool also successfully transfers execution awareness to code understanding tasks. 

\bibliographystyle{ACM-Reference-Format}
\bibliography{main}

\end{document}